\documentclass[10pt,preprint]{emulateapj}
\usepackage{xcolor}

\shorttitle{2014 Demographics Survey of the SDSS-IV}
\shortauthors{The Committee on the Participation of Women in the SDSS}

\begin{document}

\title{The SDSS-IV in 2014: A Demographic Snapshot}


\author{Britt Lundgren\altaffilmark{1,2}, Karen Kinemuchi\altaffilmark{3}, Gail Zasowski\altaffilmark{2,4}, Sara Lucatello\altaffilmark{5}, Aleksandar M. Diamond-Stanic \altaffilmark{1,6}, Christy A. Tremonti\altaffilmark{1}, Adam D. Myers \altaffilmark{7}, Alfonso Arag\'on-Salamanca\altaffilmark{8}, Bruce Gillespie\altaffilmark{4}, Shirley Ho\altaffilmark{9}, John S. Gallagher\altaffilmark{1}}

\altaffiltext{1}{Department of Astronomy, University of Wisconsin - Madison, Madison, WI 53706, USA}
\altaffiltext{2}{NSF Astronomy \& Astrophysics Postdoctoral Fellow}
\altaffiltext{3}{Apache Point Observatory/New Mexico State University, Sunspot, NM 88349, USA}
\altaffiltext{4}{Johns Hopkins University, Baltimore, MD, 21218, USA}
\altaffiltext{5}{INAF - Osservatorio astronomico di Padova, vicolo dell'Osservatorio 5, 35122 Padova, Italy}
\altaffiltext{6}{Grainger Postdoctoral Fellow}
\altaffiltext{7}{Department of Physics and Astronomy, University of Wyoming, Laramie, WY 82071, USA}
\altaffiltext{8}{School of Physics, University of Nottingham, Nottingham NG7 2RD, UK}
\altaffiltext{9}{McWilliams Center for Cosmology, Department of Physics, Carnegie Mellon University, 5000 Forbes Ave., Pittsburgh, PA 15213}

\begin{abstract}

Many astronomers now participate in large international scientific collaborations, and it is important to examine whether these structures foster a healthy scientific climate that is inclusive and diverse.  The Committee on the Participation of Women in the Sloan Digital Sky Survey (CPWS) was formed to evaluate the demographics and gender climate within SDSS-IV, one of the largest and most geographically distributed astronomical collaborations.  In April 2014, the CPWS administered a voluntary demographic survey to establish a baseline for the incipient SDSS-IV, which began observations in July 2014.  We received responses from 250 participants (46\% of the active membership).  Half of the survey respondents were located in the United States or Canada and 30\% were based in Europe.  Approximately 65\% were faculty or research scientists and 31\% were postdocs or graduate students.  Eleven percent of survey respondents considered themselves to be an ethnic minority at their current institution.  Twenty-five percent of the SDSS-IV collaboration members are women, a fraction that is consistent with the US astronomical community, but substantially higher than the fraction of women in the International Astronomical Union (16\%).  Approximately equal fractions of men and women report holding positions of leadership in the collaboration.  When binned by academic age and career level, men and women also assume leadership roles at approximately equal rates, in a way that increases steadily for both genders with increasing seniority.  In this sense, SDSS-IV has been successful in recruiting leaders that are representative of the collaboration.  That said, it is clear that more progress needs to be made towards achieving gender balance and increasing diversity in the field of astronomy, and there is still room for improvement in the membership and leadership of SDSS-IV.  For example, at the highest level of SDSS-IV leadership, women disproportionately assume roles related to education and public outreach.  The goal of the CPWS is to use these initial data to establish a baseline for tracking demographics over time as we work to assess and improve the climate of SDSS-IV.

\end{abstract}


\section{Introduction}

Many research areas in observational astronomy are moving towards a model of vast surveys driven by large national and international collaborations. In part, this migration is due to the expense of conducting experiments of increasing scope and complexity, and certainly shared capital costs are a consideration in assembling large collaborations. But beyond equipment, shared knowledge and ethos are becoming increasingly critical for efficiently conducting science. 

The Sloan Digital Sky Survey (SDSS) is a large, international astronomical collaboration
of scientists, educational personnel, engineers, computer software
and technical staff, postdoctoral associates, and students.
Planning for the original SDSS was initiated in the late 1980s by members of the Astrophysical Research Consortium, and the SDSS survey saw first light in April 2000.  Now in its fourth generation (SDSS-IV), the SDSS has involved astronomers from over 50 institutions in the US and around the world.  The SDSS is one of the longest-standing large astronomical collaborations of its size,
and its management style and policies have influenced subsequent groups,
many of which (e.g., the Dark Energy Survey) involve current and former SDSS members.

To maximize the effectiveness of any group of people working towards a goal,
the leaders and members of the group must understand the individuals involved and
their backgrounds and talents, as well as be aware of potential biases and
other factors influencing personnel decisions.  
The Sloan Foundation, a leader in funding scientific and technological research and a major contributor to SDSS operations, noted the small number of women in SDSS leadership positions during a recent survey review.   With the approval of additional funding for the SDSS-IV in 2012, the Sloan Foundation requested that the SDSS management report on efforts to recruit women into the project leadership.

In answer to this call, the SDSS leadership formed a committee ---
the Committee on the Participation of Women in SDSS (CPWS) --- and
charged it with (i) examining the ways in which collaboration
leadership is established, (ii) evaluating the diversity and general
climate within the collaboration, with a particular emphasis on gender
balance, (iii) fielding climate-related concerns from people within
the SDSS, and (iv) making recommendations to the SDSS management on
how to improve the overall equity and climate within the
collaboration.  One of the recent steps taken by the CPWS to address
this charge was to administer a collaboration-wide, voluntary
demographic survey, to establish a baseline for the incipient SDSS-IV
and to help answer the question ``Who is the SDSS?''.  The CPWS also
interviewed several members of the SDSS leadership to gain a sense of
how the recent and current leaders came into those positions.

In this report, we present the results from these surveys and
interviews.  In \S\ref{sec:policies} we describe the current SDSS
leadership recruitment policies. In \S\ref{sec:description}, we describe the creation of
the survey and present the questions asked, and in \S\ref{sec:results},
we present the resulting SDSS demographics in both textual and
graphical form.  \S\ref{sec:hiring} contains summaries of the results
of the interviews regarding leadership and recruitment, and in
\S\ref{sec:conclusions}, we draw some brief conclusions and lay out
future activities for the CPWS.  Throughout this document, we will
focus nearly exclusively on those parts of our activities concerned
with gender balance and climate, in keeping with our charge from the
SDSS leadership.  However, we note that other types of diversity ---
be it race, ethnicity, sexual orientation, age --- and a general
climate of inclusiveness and supportiveness promote the health and
productivity of the collaboration as a whole.

The SDSS collaboration includes nearly 10 percent of the professional astronomers worldwide, and the data acquired by the survey are utilized by an even larger proportion of astronomers.  As such, we hope this report will be of interest to many astronomers, both within and outside of the SDSS, and provide a quantitative basis for reflection and action with regard to the demographic balance in our professional community's largest collaborations.

\section{Previous Recommendations and Current Policies}
\label{sec:policies}

The CPWS became a standing committee in SDSS-IV in December 2013.  The
work of the current committee builds on previous recommendations from
an earlier Committee on the Participation of Women in SDSS that
existed between October 2012 and June 2013 during SDSS-III.  This
previous committee produced a report that included recommendations for
immediate and longer term actions towards providing a larger and more
representative pool of candidates for SDSS leadership positions.  The
immediate actions from this June 2013 report included the following:
\begin{itemize}
\item Treat filling SDSS management positions as akin to formal
  hiring, including active recruitment and an active announcement
  process.  This involves developing, implementing, and following a
  set of policies and procedures that are consistent with best
  inclusive practices.
\item Develop, maintain, and distribute a clear description of
  leadership roles within the collaboration.
\item Establish clear expectations for rotation of leadership
  positions and make sure people are in place to rotate into
  positions.
\item Create a standing Committee to continue to be mindful of issues
  relating to gender and participation and to implement these
  recommendations.
\end{itemize}

The June 2013 report described the procedures necessary to implement these immediate actions, and this led to an update of the official SDSS-IV policy for leadership recruitment practices in September 2013.  Quoting from the relevant SDSS-IV Practices Document, ``The primary purpose of these practices is to ensure an open process that results in project management which is representative of the collaboration, in particular in terms of gender balance.''  

Our main goal with this document is to report on the demographics of the SDSS and its leadership shortly after these new recruitment practices were implemented.  Future follow-up surveys of the SDSS, which will be conducted annually, will enable the tracking of changes in the demographics with time as well as the identification of successful policies and best-practices.

\section{Description of the Demographic Survey}
\label{sec:description}

The goals of this survey, at the start of SDSS-IV, were twofold: to assess the demographics of the collaboration across multiple parameters,
and to look for empirical correlations between those parameters and the presence of leadership responsibilities. These results also provide an important baseline for tracking changes in the makeup of the SDSS-IV over time. Determining the root of any possible correlations revealed by this survey (e.g., between leadership and gender) is a more complex undertaking.  However, a detailed study of the underlying causes of demographic trends revealed by our survey remains largely beyond the scope of this report.

Beyond measuring the overall demographics of the SDSS-IV membership, the survey was also designed to track the demographics of the collaboration leadership.  We were interested in identifying all individuals with leadership responsibilities, even if they were not officially recognized with an official title within the collaboration hierarchy (i.e., the SDSS organizational chart). Therefore, we presented members with the following definition of leadership role at the outset
of the demographics survey:
\begin{quote}
Any role whose tasks or responsibilities include making decisions that affect
other people and the survey, organizing regular project discussions or meetings, 
professional mentoring, or influencing/directing others in their tasks.
\end{quote} 
Respondents then identified whether they considered 
themselves to be ``leaders'' according to this definition, at any level --- overall SDSS collaboration, committees, and/or within
any of the individual projects (i.e., eBOSS, APOGEE, MaNGA).

\begin{figure*}
\centering
\begin{center}
\includegraphics[width=5in]{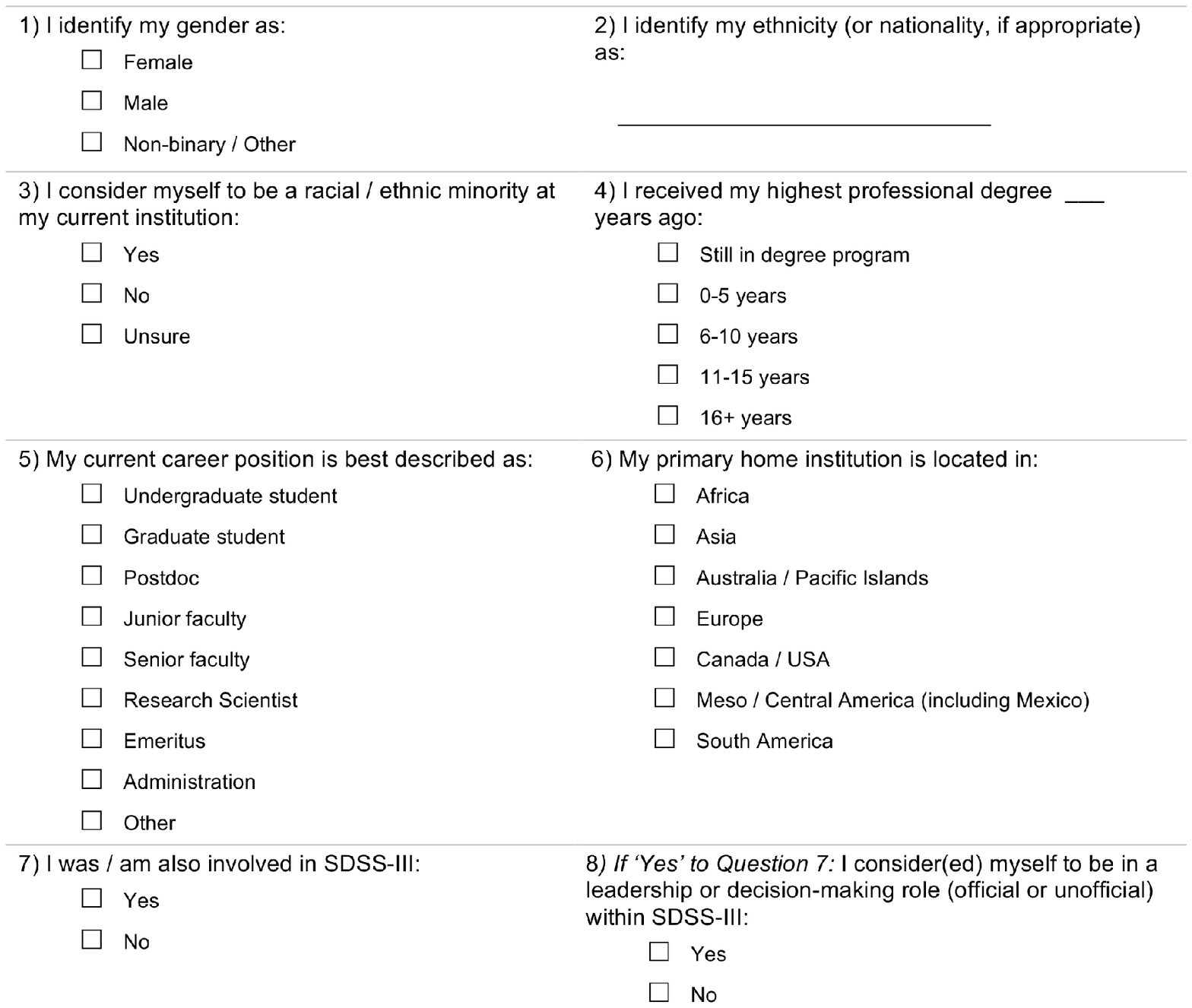}
\includegraphics[width=5in]{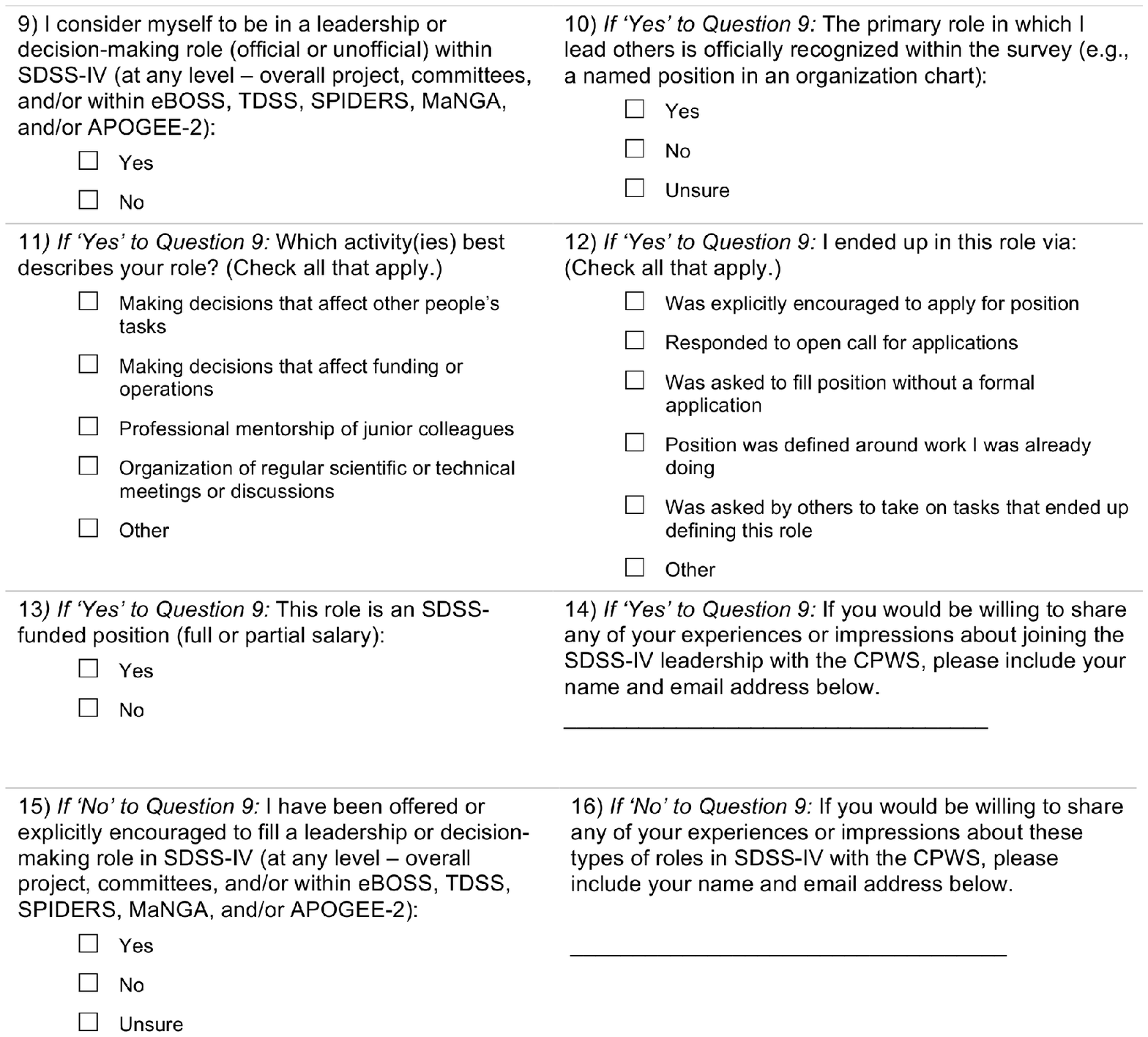}
\caption{Information requested in the SDSS-IV demographics survey.  All responses were optional. \label{fig:survey_summary}}
\end{center}
\end{figure*}

A complete list of the survey questions is provided in Figure~\ref{fig:survey_summary}.  Responding to all of the questions was optional for survey submission.  Upon completing the survey, all participants were asked if they would be willing to share any additional experiences or impressions regarding the collaboration and its leadership with the CPWS and if so, to provide their name and email address.  This question offered a straightforward option for anyone who was willing to be contacted for follow-up interviews.  

\begin{figure*}
\centering
\begin{center}
\includegraphics[width=6in]{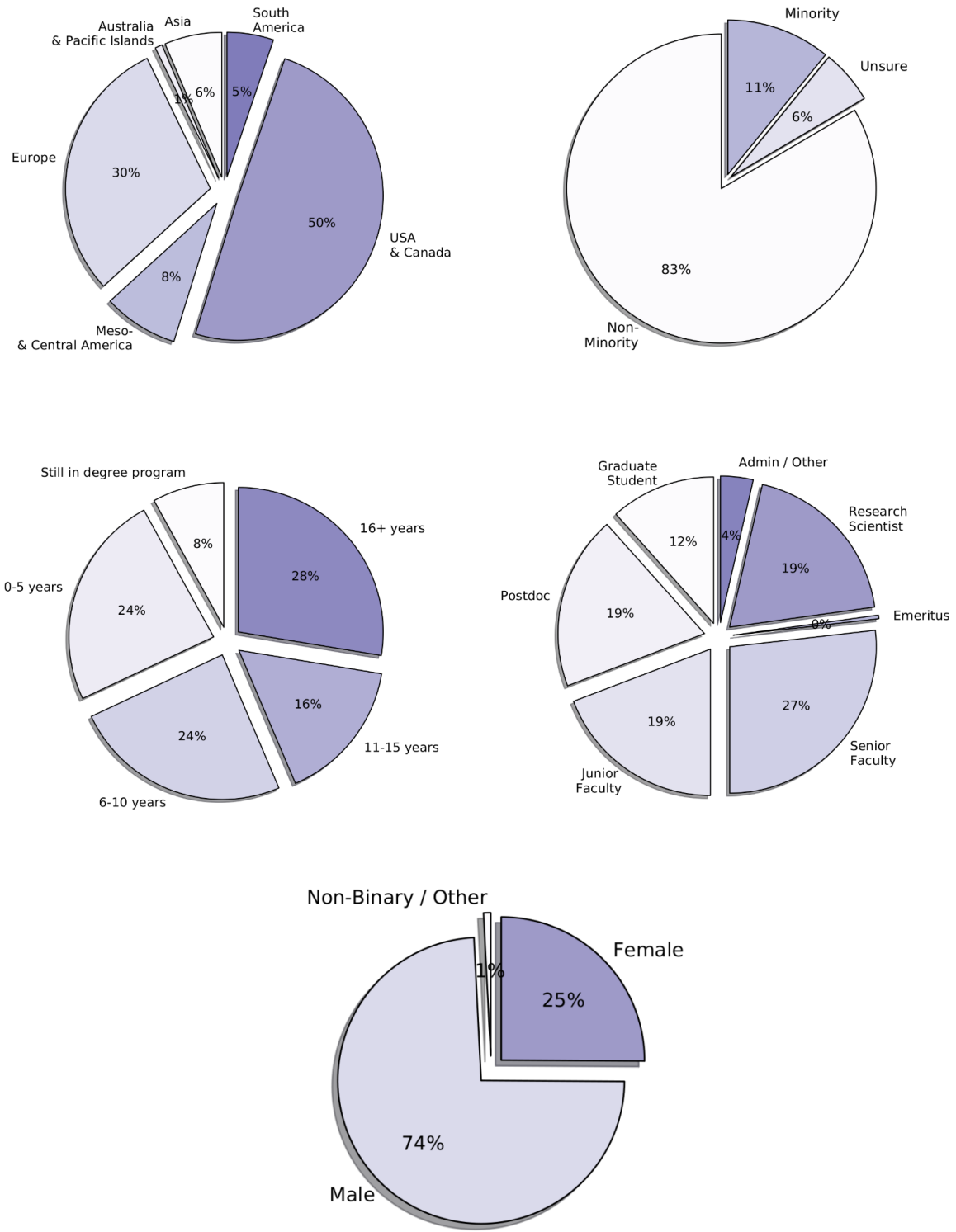}
\caption{The demographics of the survey respondents, shown clockwise from the top left in terms of geographical location, ethnic minority status with respect to one's current institution, position, gender, and years since obtaining a terminal degree. \label{fig:demographics_pie}}
\end{center}
\end{figure*}

\begin{table*}[!ht]
\centering
\begin{minipage}{0.95\textwidth}
\begin{center}
\caption{Approximate Survey Response Rate by Geographic Region \\
(Determined from 526 wiki subscribers with unambiguously reported home institutions) \label{tbl-response_rate}}
\begin{tabular}{lccc}
\hline
Geographic Region & Percent of & Survey & Percent of\\
 & Active Membership & Response Rate (\%) &  Survey Respondents \\

\hline
\hline
Asia & 15 & 19 & 6 \\
Australia / Pacific Islands & 1 & 66 & 1 \\
Europe & 24 & 60 & 30 \\
Meso/Central America (including Mexico) & 5 & 77 & 8 \\
North America (Canada; USA) & 51 & 47 & 50 \\
South America & 5 & 52 & 5 \\
\\
\end{tabular}
\end{center}
\end{minipage}
\end{table*}

\begin{table*}[!ht]
\centering
\begin{minipage}{0.95\textwidth}
\begin{center}
\caption{Ethnic Distribution of SDSS-IV Members \& Leadership based in the U.S. and Canada \\
(from all respondents with unambiguously reported ethnicity) \label{tbl-ethnicity}}
\begin{tabular}{lcccc}
\hline
& \% SDSS-IV & \% SDSS-IV & \% AAS (2013\footnote{Anderson \& Ivie 2013}) & \% USA (2013\footnote{\url{http://quickfacts.census.gov/}})\\
Ethnicity & Membership & Leadership & Membership & General Population \\
& (of 100) & (of 47) & (of 1524) & (of 316M) \\
\hline
\hline
White & 80 & 89 & 84 & 63\\
Asian or Asian American & 15 & 9 & 8 & 5\\
Hispanic or Latino & 4 & 2 & 3 & 17 \\
Black or African American & 0 & 0 & 1 & 13 \\
American Indian or Alaska Native & 0 & 0 & $<1$ & 1 \\
Native Hawaiian or other Pacific Islander & 0 & 0 & $<1$ & $<1$\\ 
Other & 1 & 0 & 2 \\
\\
\end{tabular}
\end{center}
\end{minipage}
\end{table*}

\begin{table*}
\centering
\begin{minipage}{0.95\textwidth}
\begin{center}
\caption{Gender Distribution of the SDSS-IV \\
in comparison to the larger professional astronomy community \label{tbl-gender}}
\begin{tabular}{lccc}
\hline
Organization & Men & Women & Total \\
\hline
\hline
& & & \\
SDSS-IV (Members) & 186 (74\%) & 63 (25\%) & 250 \\
SDSS-IV (Leaders) & 68 (77\%) & 20 (23\%) & 88 \\
& & & \\
\hline
& & & \\
AAS (2013\footnote{Statistics drawn from the AAS demographic report of Anderson \& Ivie (2013), citing a response rate of 66\% and a respondent seniority distribution reflective of the AAS membership as a whole.}) & 1104 (73\%) & 378 (25\%) & 1512 \\
IAU (2015\footnote{http://www.iau.org/administration/membership/individual/distribution/}) & 9537 (84\%) & 1802 (16\%) & 11339 \\
\end{tabular}
\end{center}
\end{minipage}
\end{table*}

\section{Survey Results}
\label{sec:results}

The demographic survey was completed by 250 SDSS-IV members, which accounts for 46\% of the registered SDSS-IV wiki users.  
As the collaboration wiki provides the entry point to accessing proprietary SDSS-IV data, documents, and collaborative tools, its list of subscribers effectively defines the total population of SDSS-IV members at any time.  We credit the high response rate achieved by our survey to a combination of the survey's brevity (taking $<5$ minutes to complete) and multiple collaboration-wide emails circulated by the SDSS-IV Director and Spokesperson, which strongly encouraged member participation. While we strived to representatively sample the collaboration at large, it is reasonable to assume that members with a higher degree of involvement within the collaboration were more likely to respond, thus leading to an over-sampling within our survey of individuals in positions of leadership. We recommend readers to keep this in mind, as we have, when interpreting the results in the following sections.

\subsection{Overall Demographics of the SDSS-IV Collaboration}

A graphical summary of the survey response demographics is presented in  Figure~\ref{fig:demographics_pie}.   In terms of geography, we find that the vast majority of the survey respondents work in institutions based in the United States and Canada (50\%) and Europe (30\%). Meso \& Central America, South America, Asia, Australia  \& the Pacific Islands are collectively home to 20\% of the respondents. 

As the SDSS wiki registry contains home institution information for all users, we can determine the approximate survey response rates by geographic region. These results are presented in Table ~\ref{tbl-response_rate}.  While our mean response rate across the entire collaboration was 46\%, survey participation varied strongly by geographic region, ranging from 19\% in Asia to 77\% in Meso/Central America.  The surveyÕs substantial under-representation of collaboration members from Asia is something we will work to improve upon in future surveys.

Due to the fact that the SDSS is an international collaboration, we did not provide region-specific choices for ethnicity but instead asked respondents to state their own preferred term for their ethnicity or nationality, as they deemed appropriate.  When asked generally about their minority status, 11\% of respondents reported that they considered themselves to be an ethnic minority within their home institution.  While 6\% of the respondents were unsure of their minority status, the vast majority reported that they were not an ethnic minority at their institution, suggesting that the collaboration harbors limited diversity in this context.  

For researchers in the United States, ethnic diversity has become a common metric for evaluating progress toward workplace equality in institutions of all types.  Since approximately half of the SDSS-IV researchers are employed in U.S. and Canadian institutions (with the majority of institutions in this category being in the U.S.), we can examine the ethnic diversity of the SDSS-IV collaboration for this subset of the membership with respect to the most commonly tracked ethnic groups in the United States.  In Table ~\ref{tbl-ethnicity}, we present these statistics both for the general collaboration membership, as well as for its leadership.  

Shown in comparison to the latest demographic survey of the American Astronomical Society (AAS) \citep{AndersonIvie2013}, we find relatively higher participation by Asian and Asian American researchers in the SDSS-IV but similar participation by white and Hispanic/Latino researchers.  While 24 individuals chose not to provide their ethnicity, none of the remaining 100 survey respondents self-identified as Black / African American, American Indian / Alaska Native, or Native Hawaiian / Pacific Islander.  This ethnic distribution of U.S. and Canadian researchers in the SDSS-IV is generally consistent with the larger community of astronomers in the AAS.  However, both the SDSS-IV and the AAS fall far short of the diversity (particularly with respect to Black / African American and Hispanic / Latino populations) seen in the general population of the U.S., which we present for comparison in Table~\ref{tbl-ethnicity}.  

\begin{figure*}[!ht]
\centering
\begin{center}
\includegraphics[width=5in]{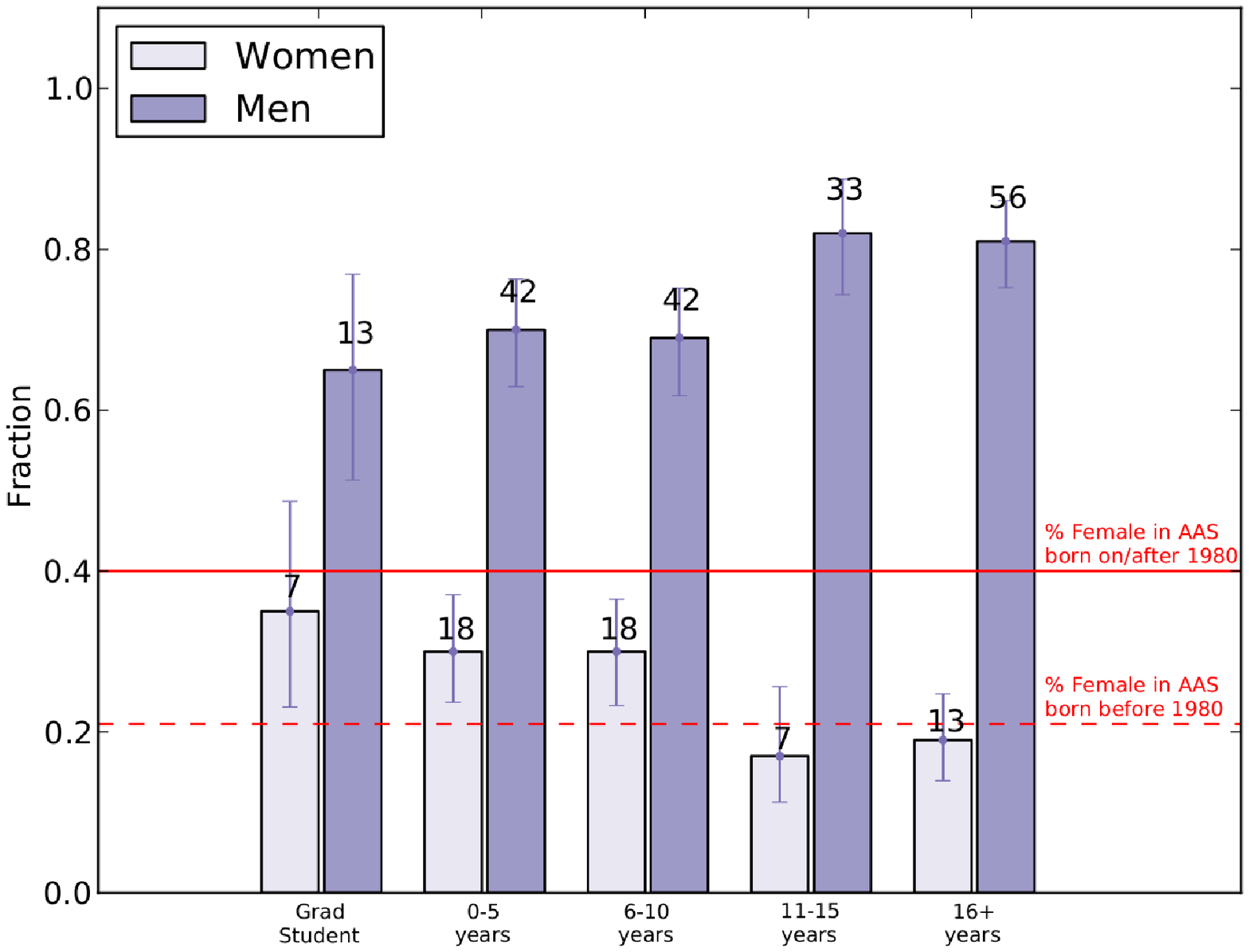}
\caption{The gender distribution of the survey respondents, shown as a function of the number of years since receipt of a terminal degree. Each pair of bars represents the fraction of respondents of each gender in that particular bin.  Generally, we find a decreasing female participation rate with increasing academic age.  To provide context, we overlay the most recent statistics for the American Astronomical Society (AAS) membership \citep{AndersonIvie2013}, which indicates a similar and well-established trend.  In this figure, and in subsequent figures, total counts are provided atop each category, and binomial 1$\sigma$ errors are provided.  \label{fig:gender_frac_years}}
\end{center}
\end{figure*}

In terms of gender, 74\% of the survey respondents identified as male, 25\% identified as female, and 1\% identified as non-binary/other. This distribution is identical to that of all registered SDSS-IV wiki users, and thus the collaboration membership at large.  In one comparison to the broader astronomical community, we find that the SDSS-IV gender distribution is identical to the most recent survey of members of the AAS, which reported 74 (25) [1]\% male (female) [non-binary] membership \citep{AndersonIvie2013}.  In Table~\ref{tbl-gender}, we also compare our statistics to the gender distribution of the International Astronomical Union (IAU), which reported only 16\% female membership in 2015.  We note that the variance in female participation amongst countries represented in the IAU is quite high, and that the geographic distribution of the IAU is not representative of the SDSS member institutions.  But overall, it appears the SDSS-IV has achieved a gender balance that is substantially better than the global average of professional astronomers.  However, it should not go unmentioned that a 25\% female fraction is still skewed dramatically from that of the human population. So there remains substantial room for improvement in the gender balance of the astronomical community, from which the SDSS-IV is not excepted.

\begin{figure*}
\centering
\begin{center}
\includegraphics[width=5in]{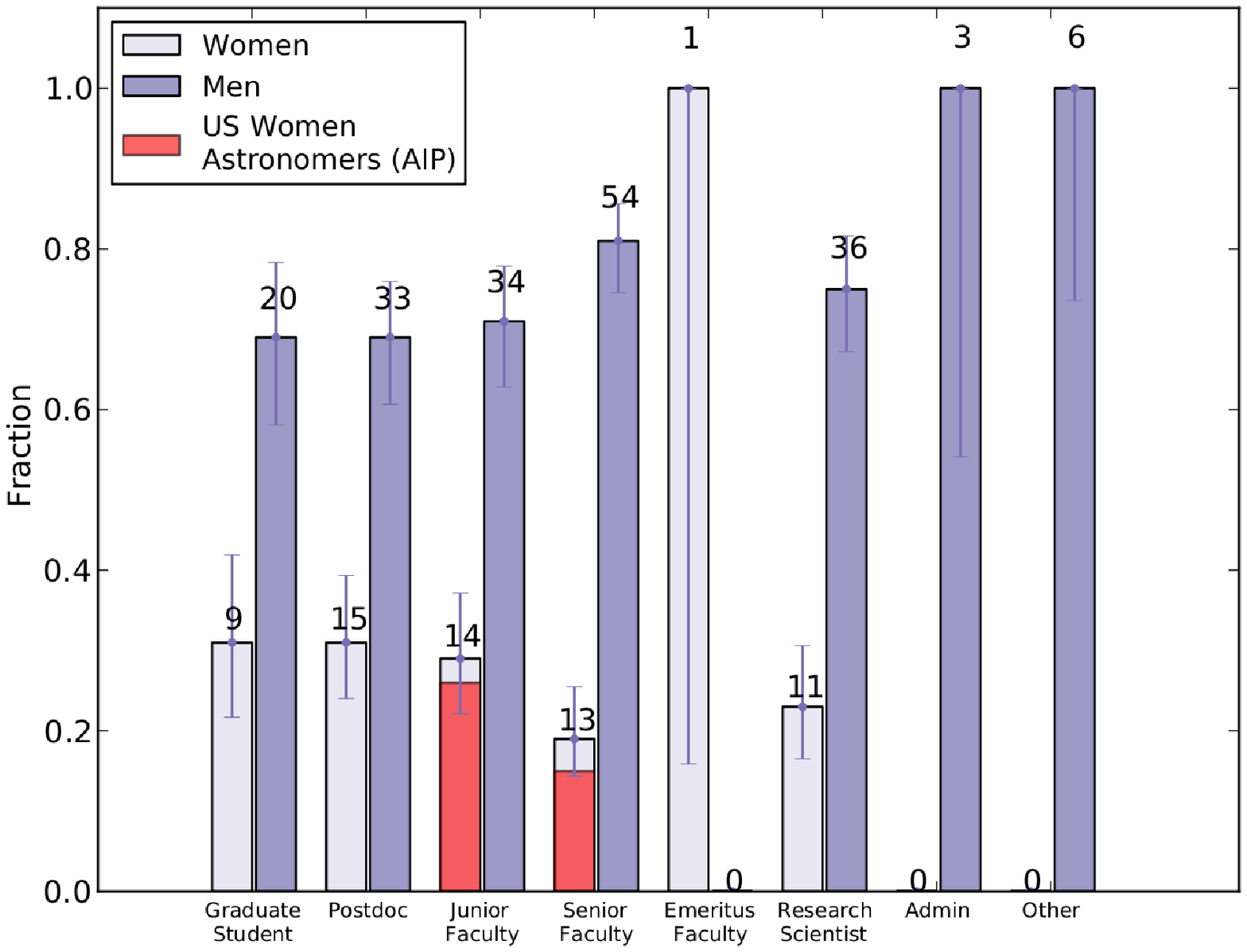}
\caption{The gender distribution of the survey respondents, shown as a function of career stage.  For comparison, we over-plot the fractions of female junior and senior astronomy faculty from a 2010 study by the American Institute of Physics \citep{Ivie2010}.  The SDSS-IV statistics appear consistent with the overall gender breakdown of astronomy faculty in the United States.\label{fig:gender_frac_position}}
\end{center}
\end{figure*}

\begin{figure*}
\centering
\begin{center}
\includegraphics[width=5in]{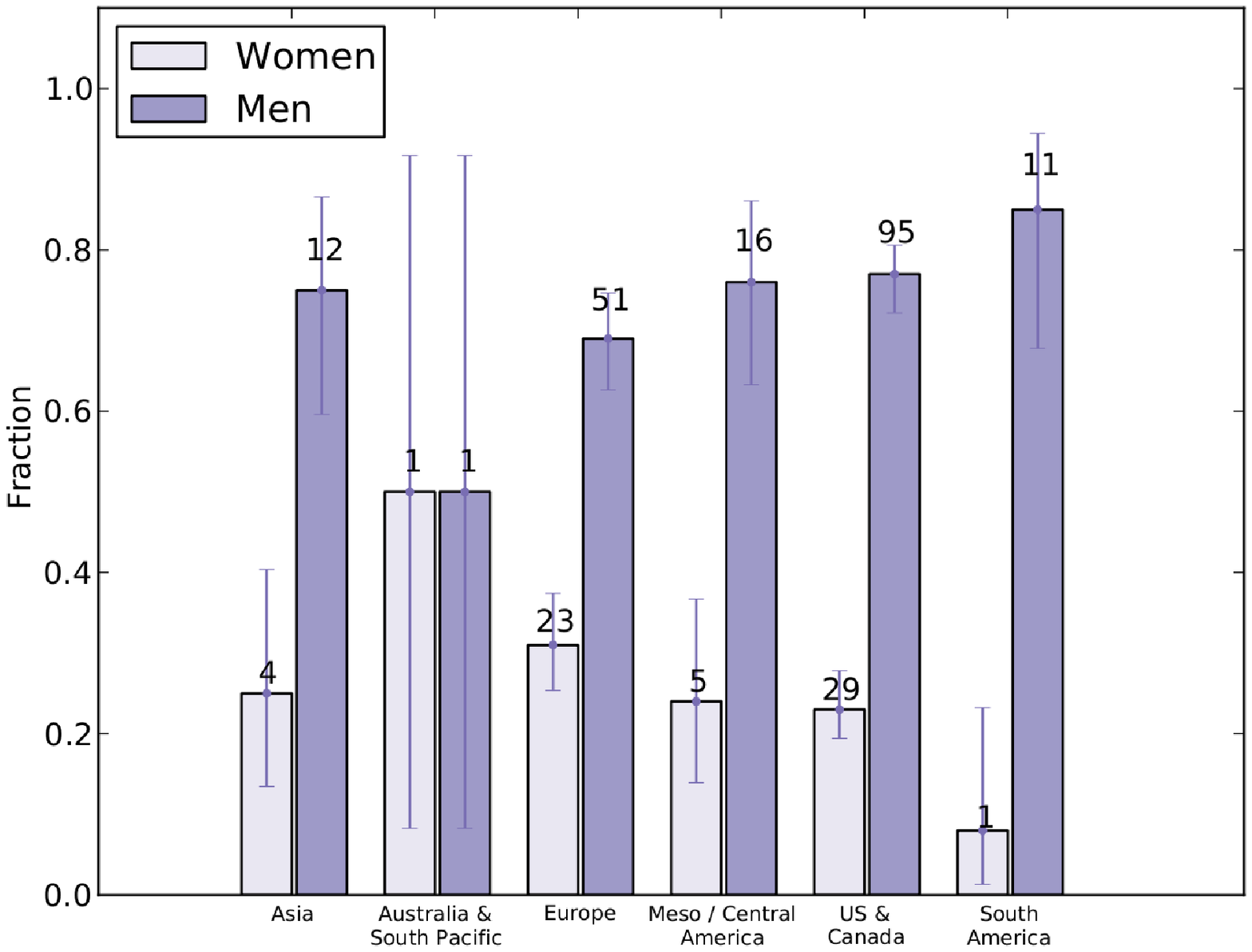}
\caption{The gender distribution of the survey respondents, shown as a function of geographic region. \label{fig:gender_frac_location}}
\end{center}
\end{figure*}

We also present in Figure~\ref{fig:demographics_pie} the distribution of the respondents broken down by the number of years since obtaining a terminal degree.  Roughly half of the survey respondents (48\%) reported having completed their terminal degree in the past 10 years.  Students represented another 8\%, with the remaining 44\% of responses coming from members more than 10 years out of their degree program.

Senior faculty comprise the most common career position amongst respondents in our survey, representing 27\% of the sample. Junior faculty, research scientists and postdocs are equally represented (each with 19\%).  Graduate students (12\%) and administrators and other positions (4\%) account for smaller shares, as shown in Figure~\ref{fig:demographics_pie}.  To clarify the discrepancy between the 12\% graduate students and the 8\% ``still in a degree program'', the respondents may have selected the ``0-5 years'' option to mean a ``student''.  Therefore, the number of people who identified themselves as a ``student'' does not equal the number of people who consider themselves ``still in a  degree program''.  For future surveys, we will better define these options.

\subsubsection{Gender balance as a function of academic age, career position, and geographic region}

As the demographics in the astronomy community are known to be changing with time, it is interesting to examine the gender breakdown of the SDSS-IV in the context of ``academic age'' and career position of our survey respondents.  Figure~\ref{fig:gender_frac_years} shows the relative male and female fractions of the survey respondents as a function of academic age, indicated by the number of years since receipt of a terminal degree. We find that the female participation rate increases with decreasing academic age, which mirrors the growth of the number of women in the field of astronomy in the last few decades.  A similar trend is observed in the AAS membership, whose most recent statistics from \citet{AndersonIvie2013} are overlaid in the plot to contextualize our results. 

\begin{figure*}[!ht]
\begin{center}
\includegraphics[width=6in]{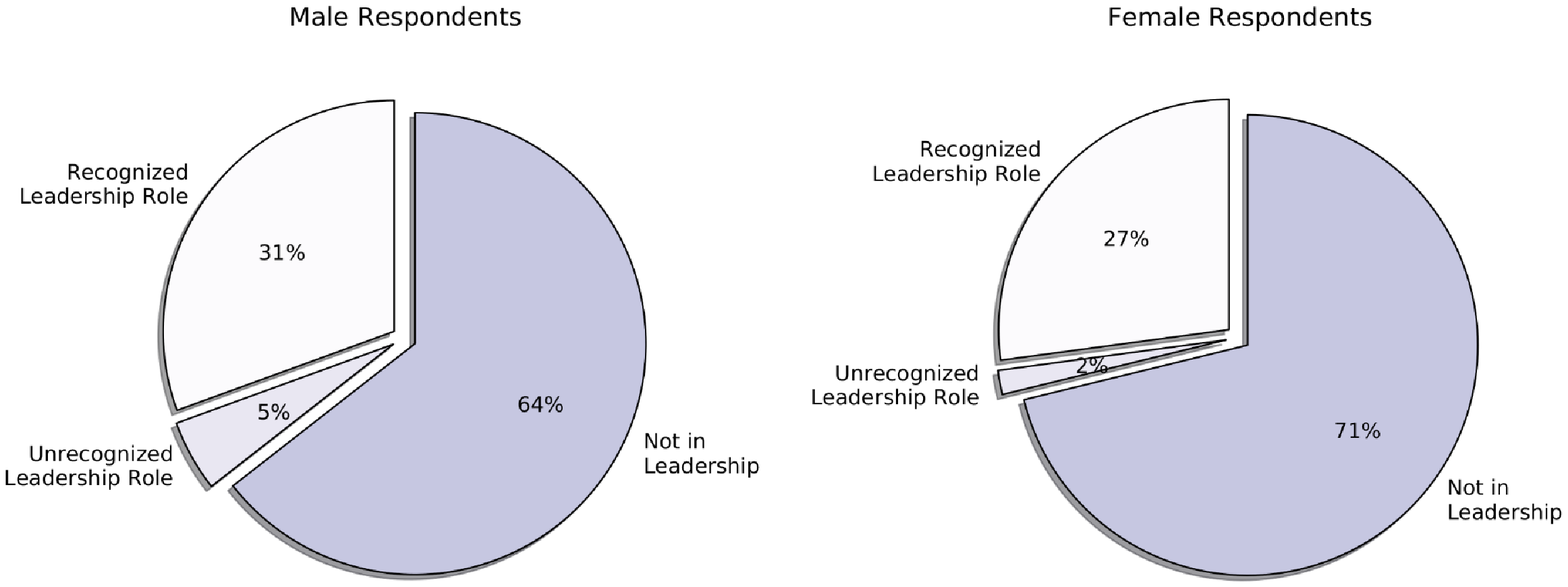}
\caption{A comparison of the fractions of male and female respondents in self-reported leadership roles.  Also shown are the fractions of responding men and women in leadership whose positions are officially recognized by the collaboration (i.e., are included in the organizational chart).  While men are slightly more likely to fill leadership roles, women holding leadership positions within the survey report higher rates of official recognition.  \label{fig:leadership_compare_pie_crop}}
\end{center}
\end{figure*}

\begin{table*}[!ht]
\centering
\begin{minipage}{0.95\textwidth}
\begin{center}
\caption{Paths to Leadership in the SDSS-IV \\ 
as Reported by 88 Survey Respondents \label{tbl-paths}}
\begin{tabular}{lccc}
\hline
Reported Path to Leadership & Men & Women & Total \\
\hline
\hline
& & & \\
Asked to fill position without a formal application & 27 (77\%) & 8 (23\%) & 35 \\
Position was defined around work I was already doing & 26 (81\%) & 6 (19\%) & 32 \\
Responded to an open call for applications & 13 (76\%) & 4 (24\%) & 17\\
Explicitly encouraged to apply for position & 12 (75\%)  &  4 (25\%) & 16\\
Asked by others to take on tasks defining the role & 10 (77\%) & 3 (23\%) & 13 \\
Other & 8 (80\%) & 2 (20\%) & 10\\
\end{tabular}
\end{center}
\end{minipage}
\end{table*}

Figure~\ref{fig:gender_frac_position} is analogous to Figure~\ref{fig:gender_frac_years} but shows the gender fractions as a function of career position.
Similarly, we observe a decrease in the number of women with career stage --- from graduate students, postdocs and junior faculty, where women comprise consistent fractions, to research scientists and even more dramatically to senior faculty. The fractions of women in junior and senior faculty positions are consistent with the gender breakdown of astronomy faculty in the U.S \citep{Ivie2010}.

Figure\ \ref{fig:gender_frac_location} presents the gender distribution of the survey respondents as a function of geographic region. 
We have small number statistics for those collaboration members reporting from Asia, Australia, South America, and Meso \& Central America.  In future demographic surveys, we hope to improve data collection from these geographical regions to get a better picture of the international collaboration.  Generally, we find no significant evidence of geographic variation in the gender ratio of SDSS-IV participants.

\subsection{Demographics of the SDSS-IV Collaboration Leadership}

In this section we examine the current leadership demographics of the SDSS-IV. While we have quantified our overall survey response rate, we do not know the bias of our responses as a function of most specific identifiers (e.g., leadership roles, career stage, minority status).   However, in comparison to the gender distribution of all active members registered on the wiki, which the CPWS could separately determine for 90\% of the registrants, we have confirmed that the gender distribution of our survey respondents is identical to that of the overall SDSS-IV membership.  Therefore, we can expect our results to be statistically unbiased at least with respect to gender It is still fair to assume that amongst the survey respondents, we are over-sampling those who are actively involved in other aspects of the collaboration and are therefore more likely to hold positions of leadership. 

In Figure~\ref{fig:leadership_compare_pie_crop} we present the overall fractions of men and women who report that they consider themselves to be in a leadership position in the SDSS-IV, as defined in Section 2.  We find that approximately equal fractions of men and women (37\% of male survey respondents and 32\% of female respondents) self-report holding positions of leadership.  Again, we stress that these numbers refer to the fraction of {\it survey respondents} and are likely over-estimates with respect to the true fraction of collaboration members in leadership positions for both genders.  Of these individuals who reported holding leadership roles, 86\% of the men and 94\% of the women described their positions as being officially recognized within the collaboration leadership structure (Figure~\ref{fig:leadership_compare_pie_crop}).  These findings suggest that women in self-reported leadership positions within the survey have slightly higher rates of official recognition, but we do not speculate on the underlying cause of this effect.

Eighty percent of survey respondents in leadership positions reported that they did not consider themselves to be an ethnic minority at their home institution.  This fraction is approximately equivalent to the demographic split of the survey respondents as a whole (80\% non-minority; 11\% minority; 9\% unsure). The ethnic distribution of the SDSS-IV leadership based in the US and Canada is presented in Table~\ref{tbl-ethnicity}.  In comparison to the SDSS-IV membership (80\% white), these leaders are slightly less diverse (89\% white).  This trend is perhaps unsurprising, given the changes in the demographics of U.S. astronomers in recent decades, combined with the fact that leadership positions draw more heavily from the senior members of the collaboration.  

In the following sections we present the demographics of SDSS-IV leadership, broken down further by academic age, career position, and organizational structure.

\subsubsection{Gender Balance in Leadership as a Function of Academic Age and Career Position}

\begin{figure*}
\centering
\begin{center}
\includegraphics[width=5in]{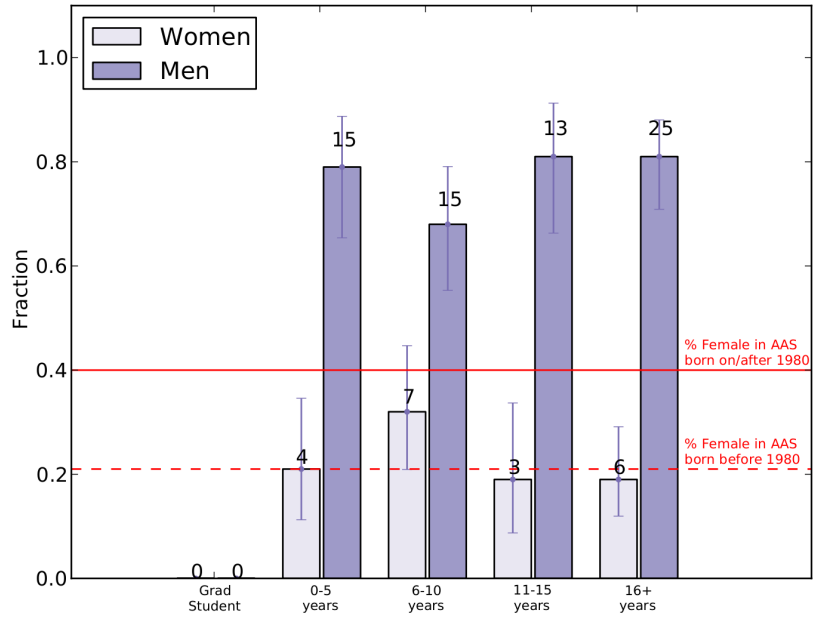}
\caption{The gender distribution of survey respondents who self-recognized as having a leadership role in the SDSS-IV, shown as a function of years since the receipt of their highest professional degree.  As in Figure~\ref{fig:gender_frac_years}, we over-plot the recent demographic results from the American Astronomical Society \citep{AndersonIvie2013}. \label{fig:leadership_gender_frac_years}}
\end{center}
\end{figure*}

In Figure~\ref{fig:leadership_gender_frac_years} we present the gender distribution of SDSS-IV members who reported holding a leadership role, shown as a function of years since obtaining a terminal degree.  We find that at all academic ages, the majority ($\sim75$\%) of leadership roles are held by men --- a fraction that is roughly consistent with the overall gender distribution of the SDSS-IV (see Figure~\ref{fig:demographics_pie}).  
As in Figure~\ref{fig:gender_frac_years}, we over-plot the recent statistics from the AAS, which indicate that the fraction of more senior members of SDSS-IV leadership who are female ($\sim19$\%) is consistent with the overall fraction of women aged 34 years and older in the AAS (21\%) \citep{AndersonIvie2013}.  However, women represent 40\% of AAS members that are younger than 34, while only 21\% of the SDSS-IV leadership positions held by collaboration members in this approximate age group are reported to be female. This may indicate a slight under-representation in SDSS-IV leadership for women in this youngest age group, but due to small number statistics, the significance of this finding is low.  We note that restricting this analysis to respondents who described their leadership roles as being ``officially recognized'' by the collaboration does not produce significantly different results.  

\begin{figure*}
\centering
\begin{center}
\includegraphics[width=5in]{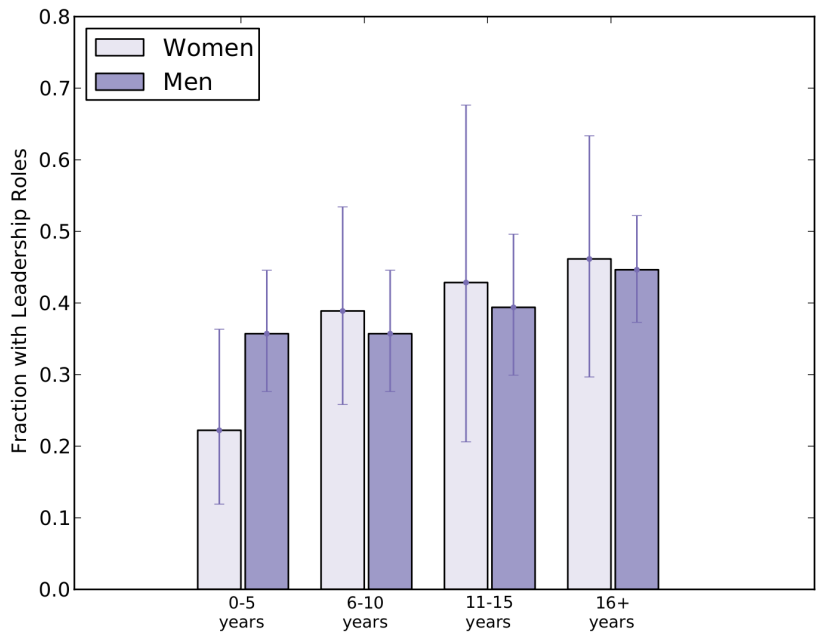}
\caption{The fraction of responding men and women in self-reported leadership roles, by years since obtaining their highest professional degree.  A steady increase in leadership responsibilities is found for both men and women as a function of academic age.  Women may be slightly less likely than their male counterparts to hold leadership roles within the first five years of completing their graduate studies, but the statistical significance of this finding is weak, and any possible discrepancy appears to rebound in later years. \label{fig:leadership_frac_years}}
\end{center}
\end{figure*}

In order to compare the rates at which men and women in the SDSS-IV assume leadership roles as a function of gender and academic age, we present in Figure~\ref{fig:leadership_frac_years} the fraction of all responding men and women in leadership, shown as a function of years since obtaining a terminal degree.  We find a statistically similar steady increase in leadership responsibilities for both men and women with increasing academic age ($\sim30$\% within 5 years of their final degree, compared to $\sim$45\% at $\geq16$ years).  Women appear slightly less likely than their male counterparts to hold leadership roles within the first five years of completing their Ph.D. ($\sim22$\%, compared to $\sim36$\%), but the statistical significance of this finding remains weak due to small numbers, and any possible discrepancy appears to rebound in later years.

\begin{figure*}
\centering
\begin{center}
\includegraphics[width=5in]{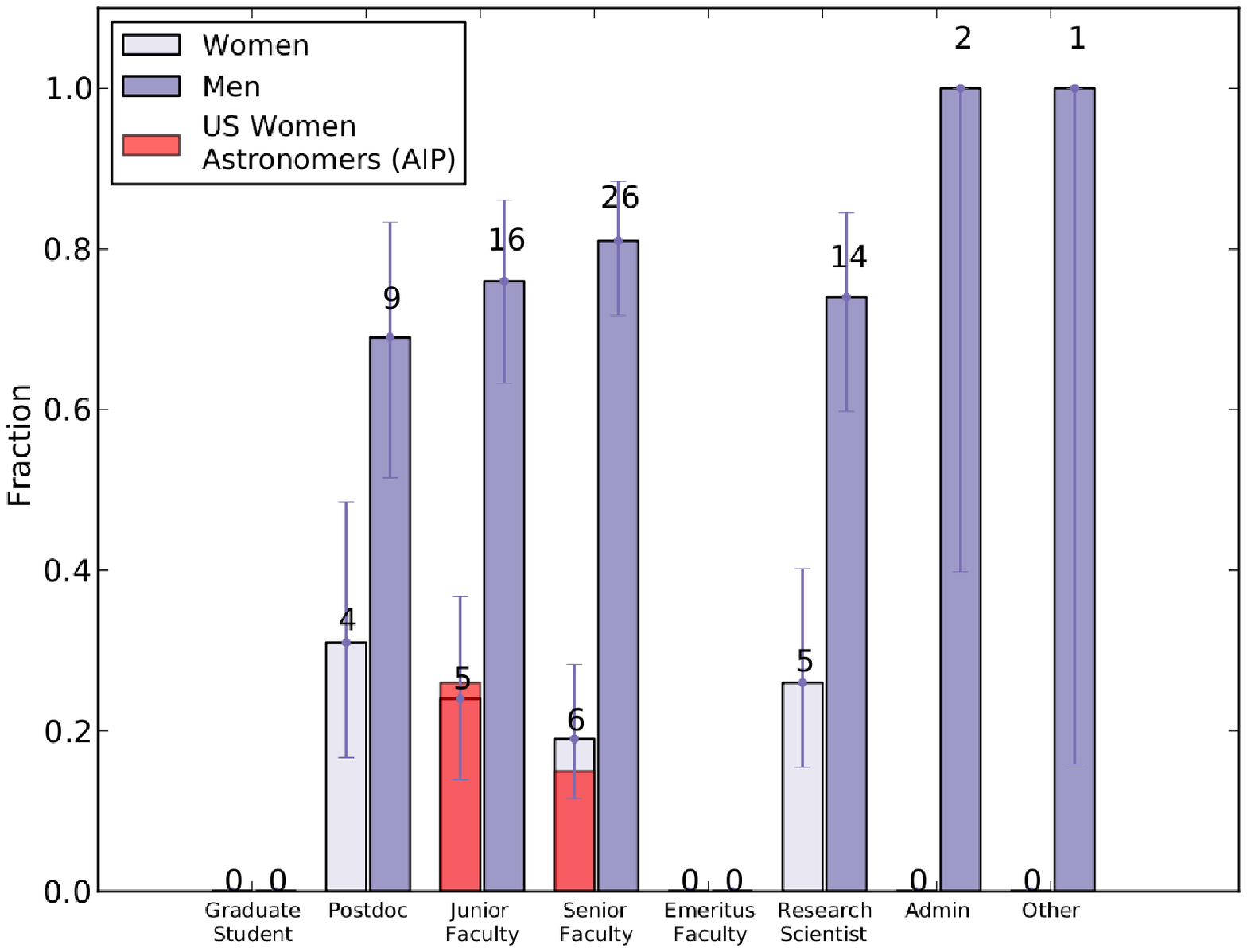}
\caption{The gender distribution of survey respondents who self-recognized as having a leadership role, shown as a function of career stage.  As in Figure~\ref{fig:gender_frac_position}, we over-plot the fractions of female junior and senior astronomy faculty in the United States from the AIP study of \citet{Ivie2010}.\label{fig:leadership_gender_frac_position}}
\end{center}
\end{figure*}

In Figure~\ref{fig:leadership_gender_frac_position} we present the gender distribution of the self-reported leaders, broken down by career position. With low significance, we find that the fraction of leaders who are female steadily decreases with increasing career level.  This trend can likely be explained by the increasing female fraction of astronomy Ph.D.s with time.  For context, we compare our data again to the recent census of junior and senior astronomy faculty by the American Institute of Physics \citep{Ivie2010}.  The fraction of astronomers at each of these career stages in the SDSS-IV leadership who are female (24\% for junior faculty; 19\% for senior faculty) is consistent with the overall gender distribution of astronomy faculty at U.S. institutions (26\%, 15\%). Again, restricting this analysis to those in officially recognized leadership roles does not produce significantly different results.

\begin{figure*}
\centering
\begin{center}
\includegraphics[width=5in]{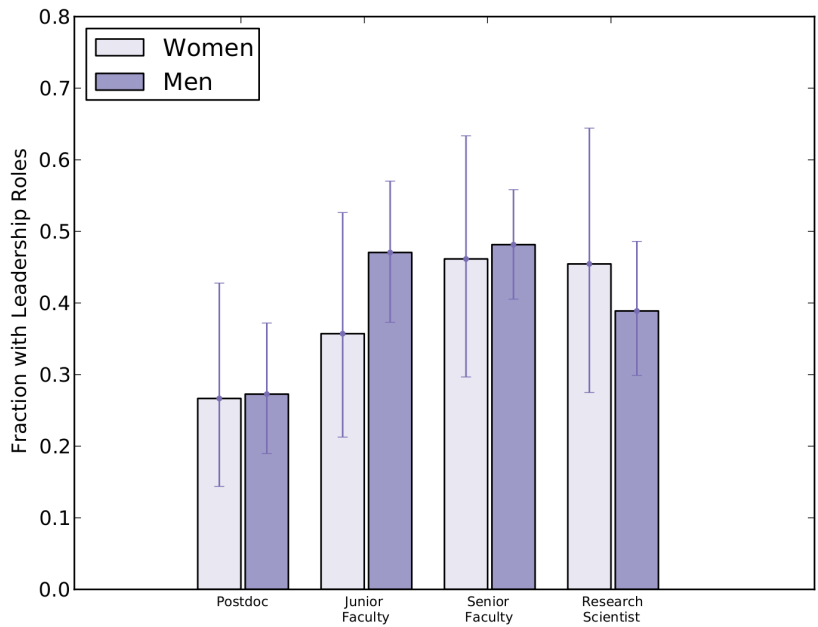}
\caption{The fraction of responding men and women in self-reported leadership roles, by position. Both genders follow a similar pattern, with the highest leadership attainment occurring amongst senior faculty. \label{fig:leadership_frac_position}}
\end{center}
\end{figure*}

In Figure\ \ref{fig:leadership_frac_position} we present the fraction of men and women in the survey who have assumed leadership roles at each career position.   We find that the fraction of both men and women respondents involved in leadership positions generally increases with career stage, from  $\sim26$\% of postdocs to $\sim48$\% of senior faculty.  Any discrepancies by gender are not statistically resolvable.

\begin{figure*}
\centering
\begin{center}
\includegraphics[width=5in]{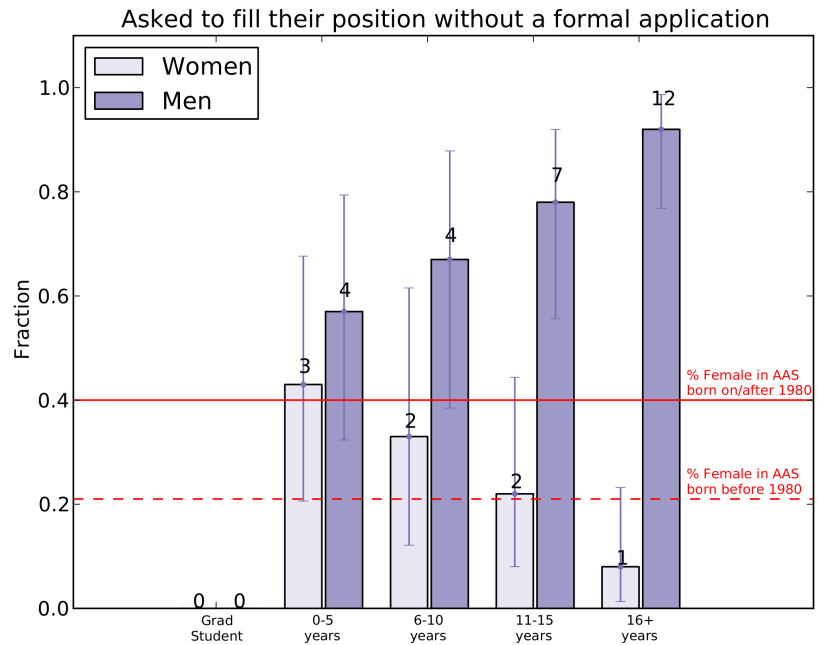}
\caption{The fraction of responding men and women in self-reported leadership roles who acquired their positions without a formal application, shown as a function of years since completing their highest professional degree. We find a strong demographic trend correlating with age, which is consistent with the changing demographics of astronomers in the U.S. \citep{AndersonIvie2013} \label{fig:leadership_gender_frac_years_noformalapp}}
\end{center}
\end{figure*}

\begin{figure*}
\centering
\begin{center}
\includegraphics[width=5in]{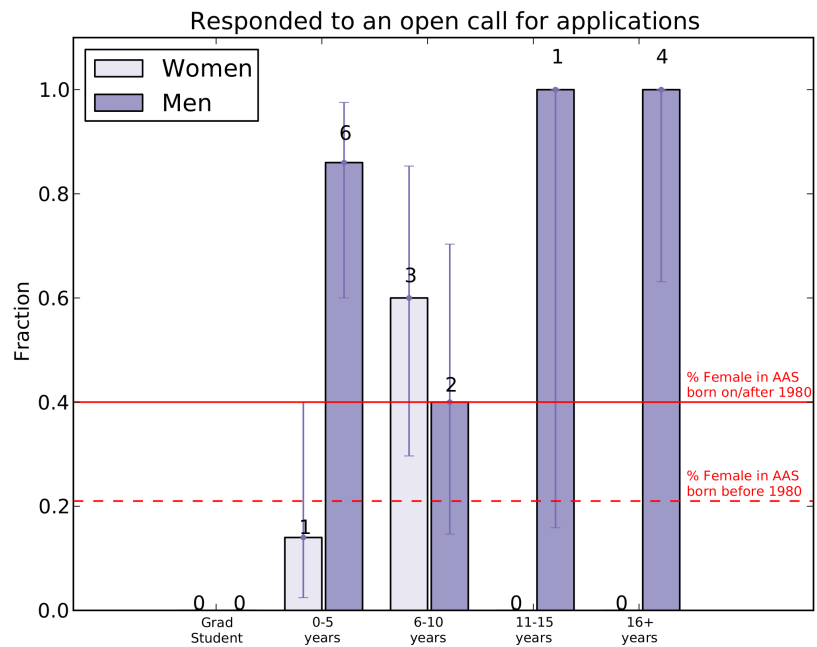}
\caption{The fraction of responding men and women in self-reported leadership roles who acquired their positions through an open call for applications, shown as a function of years since completing their highest professional degree. Despite the small numbers, we find that this path to leadership recruits preferentially from junior members of the collaboration, which tends to also have an inherently better gender balance. \label{fig:leadership_gender_frac_years_opencall}}
\end{center}
\end{figure*}

\begin{figure*}
\centering
\begin{center}
\includegraphics[width=5in]{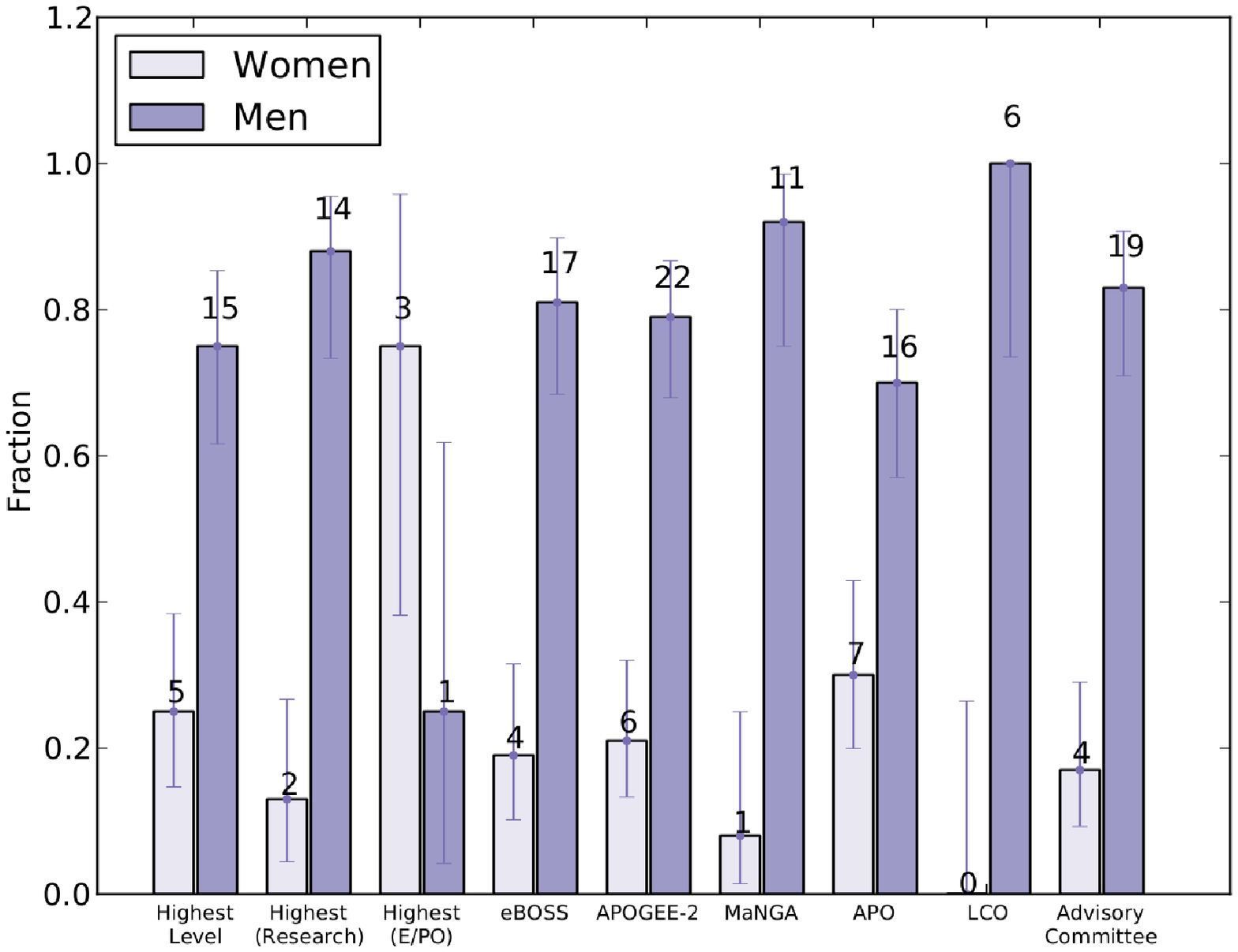}
\caption{The gender breakdown of the official SDSS-IV Collaboration organization chart, at the time of our survey. \label{fig:leaderhip_gender_frac_orgchart}}
\end{center}
\end{figure*}

\subsubsection{Self-reported Paths to Leadership}

We next consider the distribution of paths to leadership reported by survey respondents in leadership roles.  Individuals were encouraged to check all provided choices that applied to them, so the summed fractions of all responses, presented in Table~\ref{tbl-paths}, is greater than unity.  The two most commonly reported routes to SDSS-IV leadership roles were: individuals being asked to fill a position without a formal application (40\%) and having a position designed around work an individual was already doing (36\%).  Smaller fractions of respondents reported that they responded to an open call for applications (19\%) or were explicitly encouraged to apply for their positions (18\%), and 15\% reported that they were asked by others to take on tasks that eventually defined the leadership role they now hold.  An additional 11\% reported having taken some alternate path to leadership.

We find no significant differences in the overall gender balance of self-reported leaders as a function of their path to leadership (Table~\ref{tbl-paths}). In Figures~\ref{fig:leadership_gender_frac_years_noformalapp} and~\ref{fig:leadership_gender_frac_years_opencall} we examine any possible differences in the age and gender distributions of the leadership, depending on whether the positions were filled without a formal application or through an open call for applications.  Unsurprisingly, we find that positions filled without a formal application tended to be held by more senior members of the collaboration, while open calls for applications preferentially recruit younger members of the collaboration.  Interestingly, the overall gender fractions of individuals from both paths to leadership are approximately equivalent.

Approximately half (47\%) of the SDSS-IV leaders reported that they had previously held positions of leadership in the SDSS-III.  The gender ratio of this subset of the leadership is also found to be 24\% female, indicating that both new and established leaders in the SDSS-IV have a gender balance that is equivalent to that of the collaboration at large.

\subsubsection{Gender Balance in Leadership as a Function of Collaboration Structure}

In Figure\ \ref{fig:leaderhip_gender_frac_orgchart} we present the gender breakdown of the official SDSS-IV collaboration leadership, as maintained in the formal organization chart as of spring 2014.  This plot is useful for demonstrating how the leadership gender balance compares for various projects within the SDSS-IV and in comparison to the top-level leadership and management roles.  

The leadership ranks of the eBOSS, APOGEE-2, and MaNGA surveys have 19\%, 21\%, and 8\% female fractions, respectively. The Apache Point Observatory leadership is 30\% female, the Advisory Committee is 17\% female, and the Las Campanas Observatory leadership currently includes no women.  Amongst the remaining high-level leadership positions that span all surveys within the SDSS-IV collaboration, 25\% are held by women.  

Interestingly, we find that the women at these highest levels of the SDSS-IV organization disproportionately assume roles related to education and public outreach (E/PO).  Fully three quarters of the E/PO  leadership positions are held by women, compared to 12.5\% of the highest-level science-focused leadership roles.  Although the current number of E/PO roles is small (4 in total), the binomial probability that 75\% of these roles would be held by women through a random sampling of the (25\% female) collaboration membership is only 4.7\%.  Thus, we consider this finding to be statistically significant. 

The underlying cause of the over-representation of women in E/PO positions may be a complex combination of factors, which is intriguing yet beyond the scope of this report.  We do, however, note that at least one recent study has investigated the fractional participation of male and female scientists in outreach activities.  \citet{Ecklund2012} report that approximately 76\% of female physicists at top universities in the U.S. are involved in outreach activities, compared to 58\% of their male counterparts.  Thus, the over-representation of women in E/PO roles in the SDSS appears to be reflective of a larger trend in the physics and astronomy communities, at least within the U.S.

\section{Recruitment of Upper Management and Survey Leadership}
\label{sec:hiring}

In addition to the demographics survey, the CPWS separately interviewed the Principal Investigators (PIs) of the main SDSS-IV projects and the upper management.  The goal of these interviews was to provide a snapshot of the recruitment climate faced by the PIs and the leadership hierarchy within each project.  The CPWS requested the following information for each filled leadership position: a description of the position, the general qualifications the position required, the percentage of the position funded by SDSS, the career stage and gender of the person who accepted the position, the recruitment mechanism (i.e., invitation, advertisement, volunteer), the demographics of the candidate pool, and the approximate length of time of the search.

The data from these interviews indicate that many of the roles within the upper management of SDSS-IV were inherited from the previous survey management.  Often these positions were filled by people who self-nominated (and in some cases, there were no other candidates) or were invited by a committee (e.g., the ARC Board or Futures committee).  
Most of the upper management positions are held by senior faculty or research scientists.   Postdocs and junior faculty are better represented amongst the lower management.   For these positions, advertisement was the primary recruitment process.

The interviews roughly confirmed the self-reported demographic survey statistics on the various routes to leadership in the SDSS-IV.  The PIs reported that more than half of the leadership positions (56\%) were filled by individuals who either volunteered or were invited to take on the leadership position without an open search.  Such routes to leadership are often necessary in the early stages of any project due to the competitive and proprietary nature of initial proposals and funding reviews.

The PIs cited difficulties attracting a sizable pool of candidates for many leadership positions.  The workload and time commitments required by each position were commonly listed as reasons why many qualified people declined leadership positions.  In some of the lower-level management positions, the PIs report the candidate pools often had only male applicants.  Among the open searches for which the specific numbers of male and female applicants were available, we found that 21\% of the applicants were female.

Many members of the upper management noted that they were highly motivated to diversify the SDSS-IV leadership.  This objective was also emphasized to the PIs during the early stages of organizing the main SDSS-IV projects.  As described in Section 2, procedural mechanisms aimed at rectifying the collaboration leadershipÕs gender imbalance had been outlined in 2013 by the first Committee on the Participation of Women in the SDSS, in response to the Sloan FoundationÕs requests the previous year.  These policies were implemented by the PIs during the early stages of organizing the main SDSS-IV projects and may be credited to some degree with the reasonably good gender balance amongst the SDSS-IV leadership. The fact that the gender breakdown of the SDSS-IV leadership is not found to vary significantly as a function of the recruitment mechanism (Table~\ref{tbl-paths}), is interesting and could potentially indicate that the conscious prioritization of diversifying the leadership was as effective as formalizing new hiring policies.


\section{Conclusions and Future Work}
\label{sec:conclusions}

The Committee on the Participation of Women in the SDSS (CPWS) produced a demographics survey in 2014 in order to provide a snapshot of the SDSS-IV collaboration membership at the survey's outset.  Approximately 50\% of active SDSS-IV members responded to the survey.
The resulting data indicate that the overall male to female ratio of survey respondents is approximately 3 to 1.  This ratio matches the gender distribution of the SDSS-IV membership, as determined from the list of $\sim500$ SDSS-IV wiki subscribers.  A similar ratio is found for survey respondents who identified themselves as holding a leadership role within the collaboration, indicating that equal fractions of men and women are involved in positions of leadership.  

In comparison to the larger astronomical community, the 25\% female fraction in the SDSS-IV membership matches that of the 2013 US membership in the American Astronomical Society.  Given that the SDSS-IV is a large international collaboration, it is interesting to note that its female fraction is substantially higher than that of the International Astronomical Union membership (16\% in 2015).  While there is certainly much progress still to be made in obtaining better gender balance in the field of astronomy at large, the SDSS-IV collaboration appears to be doing relatively well in this regard.  We are pleased to report that the SDSS-IV has recruited its leadership in a way that mirrors the above-average gender balance of its membership.  

Approximately 11\% of survey respondents indicated that they considered themselves to be an ethnic minority at their current institution.  The ethnic breakdown of the members based in the U.S. and Canada is comparable to the current demographics of the AAS, but both measures indicate a substantial under-representation of Black, Hispanic, and Native American members in astronomy, relative to the general population in the U.S.  Nearly equal fractions of the SDSS-IV membership (83\%) and leadership (80\% ) reported that they did not consider themselves to be an ethnic minority at their current institution, indicating that the ethnic distribution of the leadership is generally consistent with that of the collaboration at large.

Of the self-identified leaders responding to the survey, $>85$\% of both men and women reported that their roles were officially recognized positions within the collaboration.  The gender balance of the SDSS-IV leadership declines with increasing academic age and career level, and this trend may be due in part to the decreasing female fraction with increasing seniority that persists within the SDSS-IV collaboration as well as the astronomy community at large.  When binned by academic age and career level, men and women assume leadership roles at approximately equal rates, in a way that increases steadily for both genders with increasing seniority.  We note that at the highest level of SDSS-IV leadership, women disproportionately assume roles related to education and public outreach.

The CPWS will continue monitoring the growth and changes of the SDSS-IV collaboration membership and leadership during the projected lifetime of SDSS-IV.  We are already expanding our investigations to include other types of diversity (e.g., sexual orientation and transgender status, disability, civil and family status) to obtain a better understanding of our international collaboration.  As the project matures and more positions are filled through open advertisements, we plan to evaluate whether the recruitment policies outlined in Section 2 prove effective in further diversifying the SDSS leadership. In a future report, we hope to present data-driven recommendations for diversifying the membership and leadership in astronomy collaborations of similar scale.  In the meantime, demographic data from other large collaborations would also be useful for determining whether the SDSS is a typical case.

\acknowledgements

Funding for the Sloan Digital Sky Survey IV has been provided by the Alfred P. Sloan Foundation and the Participating Institutions. SDSS-IV acknowledges support and resources from the Center for High-Performance Computing at the University of Utah. The SDSS web site is www.sdss.org.

SDSS-IV is managed by the Astrophysical Research Consortium for the Participating Institutions of the SDSS Collaboration including the Carnegie Institution for Science, Carnegie Mellon University, the Chilean Participation Group, Harvard-Smithsonian Center for Astrophysics, Instituto de Astrof'sica de Canarias, The Johns Hopkins University, Kavli Institute for the Physics and Mathematics of the Universe (IPMU) / University of Tokyo, Lawrence Berkeley National Laboratory, Leibniz Institut f{\"u}r Astrophysik Potsdam (AIP), Max-Planck-Institut f{\"u}r Astrophysik (MPA Garching), Max-Planck-Institut f{\"u}r Extraterrestrische Physik (MPE), Max-Planck-Institut f{\"u}r Astronomie (MPIA Heidelberg), National Astronomical Observatory of China, New Mexico State University, New York University, University of Notre Dame, Observat{\'o}rio Nacional do Brasil, The Ohio State University, Pennsylvania State University, Shanghai Astronomical Observatory, United Kingdom Participation Group, Universidad Nacional Aut{\'o}noma de M{\'e}xico, University of Arizona, University of Colorado Boulder, University of Portsmouth, University of Utah, University of Washington, University of Wisconsin, Vanderbilt University, and Yale University.



\begin{thebibliography}{}
\bibitem[Anderson \& Ivie(2013)]{AndersonIvie2013} Anderson, G., Ivie, R. 2013 ``Demographics Survey of 2013 US AAS Members", Statistical Research Center of the American Institute of Physics
\bibitem[Ecklund et al.(2012)]{Ecklund2012} Ecklund E. H., James S. A., Lincoln A. E. (2012) ``How Academic Biologists and Physicists View Science Outreach'', PLoS ONE 7(5): e36240. doi:10.1371/journal.pone.0036240
\bibitem[Ivie et al.(2010)]{Ivie2010} Ivie, R., White, S., Garrett, A., and Anderson, G. 2010 ``Women among Physics \& Astronomy Faculty: Results from the 2010 Survey of Physics Degree-Granting Departments", Statistical Research Center of the American Institute of Physics 
\end{thebibliography}
{}

\end{document}